\documentclass[aip,jap,showpacs,12pt,onecolumn,preprint]{revtex4-1}
\usepackage{graphicx}
\usepackage{bm}
\usepackage{amsfonts}
\usepackage{amssymb}
\usepackage{amsmath}
\usepackage{graphicx}
\usepackage{dcolumn}
\usepackage{bm}

\begin{document}

\title{Influence of the Surface Structure and Vibration Mode on
the Resistivity of Cu Films}
\author{Ya-Ni Zhao}
\author{Shi-Xian Qu}
%\thanks{Author to whom correspondence should be addressed}
%\email{sxqu@snnu.edu.cn}
\affiliation{Institute of Theoretical \&
Computational Physics, School of Physics and Information Technology,
Shaanxi Normal University, Xi'an 710062, China}

\author{Ke Xia}
\affiliation{Department of Physics, Beijing Normal University,
Beijing 100875, China}

\begin{abstract}

The influence of the surface structure and vibration mode on the
resistivity of Cu films and the corresponding size effect are
investigated. The temperature dependent conductivities of the films
with different surface morphologies are calculated by the algorithm
based upon the tight-binding linear muffin-tin orbital method and
the Green's function technique. The thermal effect is introduced by
setting the atomic displacements according to the Gaussian
distribution with the mean-square amplitude estimated by the Debye
model. The result shows that the surface atomic vibration
contributes significantly to the resistivity of the systems.
Comparing the conductivities for three different vibration modes, it
is suggested that freezing the surface vibration is necessary for
practical applications to reduce the resistivity induced by the
surface electron-phonon scattering.
\end{abstract}
\maketitle

\section{Introduction}
 The study on the transport property of
copper has been attracting much attentions of
researchers\cite{Werner,V.Barnat,Aouadi,J.J.Plombon,Youqi
Ke,J.M.Purswani,Tik Sun,P.X.Xu} since it often serves as a good
conductor in devices either on macro or micro scales, such as the
interconnect material of integrated circuits (IC). Reducing
resistivity is highly demanded in IC technology to cut down the
power consumption of micro-devices. Effort should be made to reach
this goal theoretically and experimentally. It was found that the
decreasing cross section of Cu wire would lead to the increase of
the resistivity\cite{J.J.Plombon,Aouadi,Youqi Ke}, which is the so
called ``size effect". It may give rise to $100\%$ increase in the
resistivity when the size of the sample is below
50~nm.\cite{J.J.Plombon,Aouadi} This effect severely impacts the
time delay of the interconnects of integrated circuits and thus
represents a major challenge for the continuing evolution of the
microelectronic devices.

Actually, the size effect is related to the increasing surface to
bulk atomic number when the dimension of the cross section of thin
films decreases to nanometer scale. Furthermore, experimental
observations revealed that there was about $-2.0\%$ contraction of
the top-layer for Cu films when $T\leq 305~{\rm K}$, and $-2.3\%$
when $T>520~{\rm K}$.\cite{D.E.Fowler} Generally, the total
resistivity of Cu interconnect origins from several scattering
mechanisms, including lattice vibrations, impurities, defects,
surface roughness and grain boundaries, \emph{etc}. Surface
scattering is considered to play a key role in the increase in the
resistivity of copper thin films.\cite{Youqi Ke,Zong}  Therefore,
investigating the influence of the surface structure and its
vibration mode on the resistivity, i.e. the effect of the
surface electron-phonon interaction, is a very important topic.

Many methods have been advanced to study the effect of the surface
electron-phonon scattering on the resistivity. A widely used
semiclassical model is the Fuchs-Sondheimer
model,\cite{Fuchs,Sondheimer} where a phenomenological parameter is
used to characterize the electron scattering at the surface. Other
more advanced analytic models\cite{Zhang1995,Palasantzas,Meyerovich}
have also been proposed in the general area of the thin-film
resistivity, which take into account the quantum effects that may
become prominent at very small film thickness. Recently, parameters
free \emph{ab initio} methods\cite{Timoshevskii,Zhou} have been used
to directly calculate the resistivity of Cu films and nanowires,
where a supercell approach was employed on periodic atomic
structures. In an earlier work\cite{Ke Xia}, we have developed an
approach based upon a tight-binding muffin-tin-orbital
implementation of the Landauer-B\"{u}ttiker formulation of transport
theory within the local-spin-density approximation of
density-functional theory. It was used to calculate the resistivity
due to diluted impurities in alloys\cite{P.X.Xu} and good agreement
with experiment\cite{Bass82} was obtained. Unfortunately the code
works only for occasions at zero temperature. In the current work,
however, we are interested in the finite temperature effect of the
surface electron-phonon scattering on the resistivity of thin films.
Therefore the approach should be revised to include the thermal
effect. It is achieved in this paper by sampling the disordered
atomic configurations due to thermal vibration in the scattering
region according to Gaussian distributions, where the temperature
dependent mean-square amplitudes are estimated by the Debye model.

The rest of this paper is organized as follows. In
Sec.~\ref{section2}, we describe the atomic structure of Cu films,
the determination of the atomic displacements due to thermal
vibration, and the computational method. The temperature dependent
resistivity of the bulk and the film structures of copper are
presented and discussed in Sec.~\ref{section3}. A short summary is
made in Sec.~\ref{section4}.

\section{Model and Computation Method}\label{section2}

\subsection{Geometry Structures}
\begin{figure}
\begin{center}
\includegraphics[width=3in]{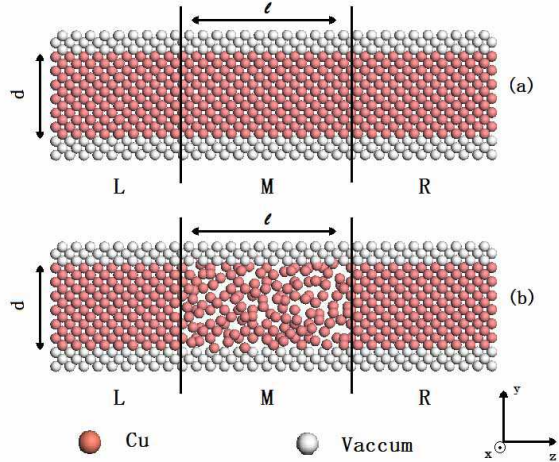}
\caption{\label{bentstraightwire figure} (Color online) The atomic
structure of Cu film consisting of two lead regions (denoted by L
and R) and one scattering region (denoted by M). (a) for $T=0~{\rm
K}$, no disorder in the scattering region, except for the quantum fluctuation
of lattice; (b) for $T\neq 0~{\rm K}$, the atoms in scattering region violate the perfect crystal
structure.}\label{geo}
\end{center}
\end{figure}

The geometry structure of the system studied in the current work is
shown in Fig.~\ref{geo}. It consists of three parts,
i.e. the two contacts regions denoted respectively by L and R, and a
scattering region denoted by M. The L and R regions are Cu metals
with perfect fcc structure, and the lattice constant $a$ is set to
$3.61~{\rm \AA}$. The $x$, $y$, and $z$ axes are along
$\langle 10\bar{1}\rangle$, $\langle 010\rangle$ and $\langle 101\rangle$ directions, respectively. $d$ is
the thickness of the Cu film.

The atomic structure is assumed to be a perfect crystal in the
scattering region when the temperature is $0~{\rm K}$ (see
Fig.~\ref{geo}(a)). For finite temperature, the atoms deviate from
their equilibrium positions due to thermal vibration (as shown in
Fig.~\ref{geo}(b)), which can be simulated by random
displacements.\cite{John H. Barrett} Actually, the temperature
dependence of the mean-square displacement has been studied both
theoretically\cite{G.Beni} and experimentally.\cite{C.J.Martin} Here
we introduce it by a simple argument based upon the Debye model.

In the language of phonon, the lattice vibrations is described by the
linear combination of collective oscillations with different frequencies. The energy of each atom contributed
by one collective mode with frequency $\omega$ can be represented as
$E=\frac{1}{2}m\omega^{2}A^{2}$ in average, which gives the square of atomic vibration amplitude
\begin{equation}
A^{2}=\frac{2E}{m\omega^{2}},\label{amplitude}
\end{equation}
with the single atom mass $m$. On the other hand, the total energy $E$ in equation (\ref{amplitude})
can be estimated from the statistical average of phonon number $\langle n\rangle$
\begin{equation}
E=(\langle n\rangle+{1\over 2})\hbar\omega,
\end{equation}
from the Planck distribution$\langle n\rangle\equiv{1\over{\exp(\hbar\omega/k_BT)-1}}$.

We further assume that the frequency distribution of the collective
oscillations obeys the Debye model, and thus the mean-square amplitude can be expressed by
\begin{equation*}
\langle A^2\rangle={2\over m}{\int^{\omega_D}_{0}D(\omega){E\over
\omega^2}d\omega \over{\int^{\omega_D}_{0}D(\omega)d\omega}},
\end{equation*}
where $\omega_{D}$ is the Debye frequency, and $D(\omega)$ is the
density of state for phonons. For our case, we have
$D(\omega)={3\omega^2/{2\pi^2 v^3}}$, where $v$ is the sound
velocity in the system. Finally, we get the analytic expression for
the mean-square amplitude, i.e.
\begin{equation}
\langle A^2\rangle={6\hbar^2\over {mk_{B}\theta_D}}[{1\over 4}+({T\over
\theta_D})^2\Phi_1]\label{amplitude1},
\end{equation}
where $\theta_{D}$ is the Deybe temperature, and $\Phi_n\equiv\int^{\theta_D/T}_0 dx{x^n\over{e^x-1}}$.

Now, the mean-square vibrational amplitudes $\langle A^{2}\rangle^{1/2}$ of the
system at finite temperature can be calculated through equation
(\ref{amplitude1}), and the atomic positions in the scattering
region in Fig.~\ref{geo}(b) can thus be determined through the
random displacements satisfying the Gaussian distribution
$P(x)=(2\pi u^2)^{-1/2}e^{\frac{1}{2}x^2/{u^2}}$, with the
mean-square displacement $u^2=\langle A^2\rangle/2$. Here, our approach to describe the lattice vibration
is more suitable for weakly correlated atoms, and is a rough approximation for the calculations blow.

\subsection{Computational Method}
The transport property of the system is estimated by the first
principle method developed in the previous works.\cite{Youqi Ke,Ke
Xia} In this approach, the electron structure is calculated through
the tight-binding linear muffin-tin orbital method, where the atomic
sphere potentials in scattering region are determined by the
self-consistent calculations based upon the Green's function
technique. The conductance is calculated by the
Landauer-B\"{u}ttiker formulism, and the resistivity is
expressed explicitly in terms of the transmission matrix $T$ as\cite{schep}
\begin{equation}
\rho\equiv {S R_{L/R}\over \ell}={S\over \ell}{h\over e^2}[\frac{1}{\sum T_{\mu\nu}}-\frac{1}{2} (\frac{1}{N_{L}}+\frac{1}{N_{R}})].\label{schepform}
\end{equation}
Here, $S$ is the interface area and $\ell$ is the length of the
lateral supercell, which is used to model the temperature induced
disorder. The element $T_{\mu\nu}$ is the transition probability from
state $|\nu\rangle$ in the left-hand lead to state $|\mu\rangle$ in the right-hand lead,
and $N_{L}(N_{R})$ is the number of conduction channels in lead L(R), which gives the Sharvin conductance\cite{schep}.
The transmission matrix can be obtained by the wave function matching at the interfaces between
the leads and the scattering region defined in the previous
subsection. In the numerical realization, to improve the accuracy,
the resistivity at a given temperature is determined by the linear
regression of the resistances obtained for different lengths.

The computational method described above is in the ballistic
transport region, where there is no energy loss for electrons.
However, the electron-phonon scattering should be an inelastic
scattering processes. Thus one may question whether it is valid to
describe the inelastic process by the current approach. This has in
fact been discussed in the previous work in detail.\cite{T.A.Abtew}
The authors proved that the average conductivity over time-dependent
configurations will give the same results as the static
electron-phonon scattering method does. In the current work, we
carry out the average of the conductivities obtained for different
atomic configurations determined by the Gaussian distributions with
different random number sequences.

\section{Results and Discussion}\label{section3}
\subsection{Bulk Cu}
To test the validity of the approach in the current work, we
calculate the temperature dependent resistivity of bulk copper for
test purpose firstly. The Debye temperature in
Eq.~(\ref{amplitude1}) is set to $315~{\rm K}$ in the
calculations\cite{Donald S. Gemmell}. The theoretically estimated
and the experimentally measured\cite{C.J.Martin} mean-square displacements of atomic vibration for different
temperatures are listed in Table \ref{tableampl} for comparison. It
is shown that the formers are at most $9\%$ less than the
latters due to the anharmonicity of the lattice potentials.\cite{D.E.Fowler} Therefore we can use the theoretically estimated
mean-square displacements in the current work.

\begin{table}[!h]
\tabcolsep 5mm \caption{The theoretically estimated and experimentally measured values of the
mean-square atomic vibration displacements for different temperatures $T$ with $\theta_{D}=315~{\rm K}$.}\label{tableampl}
\begin{center}
\def\temptablewidth{0.42\textwidth}
\begin{ruledtabular}
\begin{tabular}{ccc}
{$T(K)$}&{Theoretical Estimations $({\rm
\AA})$}&{Experimental Measurements\cite{C.J.Martin} $({\rm
\AA})$}
\\ \hline
107  &  0.055   & 0.055     \\
305  &  0.086   & 0.085     \\
520  &  0.111   & 0.115     \\
685  &  0.127   & 0.134     \\
900  &  0.146   & 0.160     \\
\end{tabular}
\end{ruledtabular}
\end{center}
\end{table}

\begin{figure}
\includegraphics[width=3.5in]{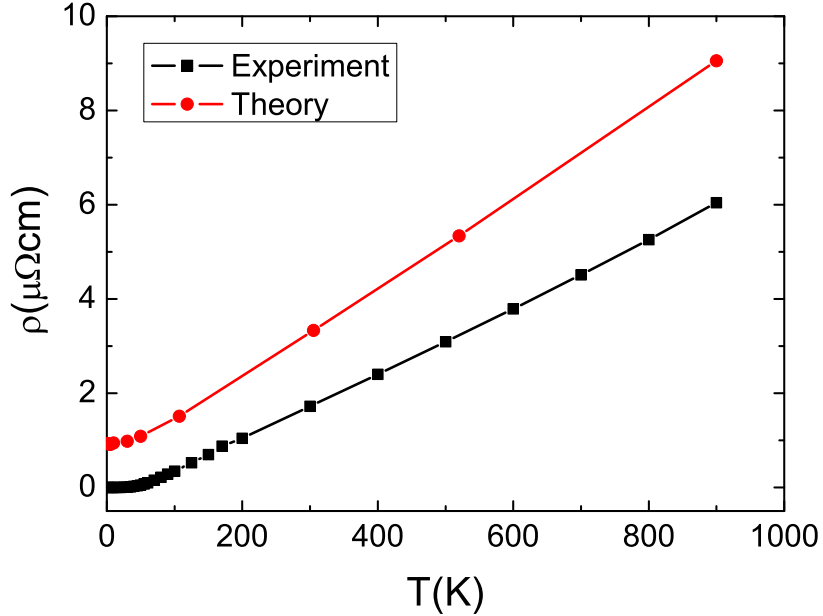}
\caption{\label{bentstraightwire figure} (Color online) Resistivity
as a function of temperature. The theoretical results are show as
red line, The experimental results are show as black
line.}\label{bulkcu}
\end{figure}

For the bulk Cu structures, the transport is along the $\langle 001\rangle$
direction, and the lateral supercell is $10\times10$ in size. The
calculated resistivity is shown in Fig.~\ref{bulkcu} (see the red
dot line). One may find that it increases linearly with the
temperature when T is above $100~{\rm K}$, and the slope is
$9.32\times10^{-3}~{\rm \mu}\Omega {\rm cm/K}$. The experimentally
measured resistivity\cite{P. A. Matula} (the black dot line in
Fig.~\ref{bulkcu}) shows a similar behavior to the calculated one,
but with a 25\% smaller slope, which is $7.02\times10^{-3}~{\rm
\mu}\Omega {\rm cm/K}$. We have to point out that the total resistivity
no longer shows the linear relation at low temperature region, which
is due to the fact that the energy equipartition theorem fails in
this region.

One may notice that the calculated values of the resistivity are
larger than the experimental ones, and is not zero at the zero-temperature as expected.
This is understood from the atom vibration amplitude estimated by relation (\ref{amplitude1}),
which gives a non-zero vibration at zero-temperature. This comes from the zero-point energy of
the phonons, and the non-zero resistivity at zero-temperature is from the quantum fluctuation of
the crystal lattice\cite{R. L. Libo}. However the close slope of temperature dependence of the resistivity between
the theoretical and experimental results suggests that the method employed in this
work is able to capture the main physical picture of the electron-phonon scattering.

\subsection{Cu Film}
As shown in the previous works,\cite{Youqi Ke} the electrical
resistivity is largely dependent on the surface scattering when
materials are in nanometer scale. Thus we expect the significant
contribution from the surface electron-phonon scattering to the
resistivity in nanostructures. In this subsection, we calculate the
temperature-dependent resistivity of thin films in the same way as
for the bulk samples. The films with different thicknesses $d$,
denoted by the number of monolayers (ML), are considered to study
the size effect. In the scattering region, the lateral supercells is
used to deal with the thermal vibrations, similar to the bulk copper.

We treat two outer MLs of the Cu films as its surface, and the other
inner MLs as its "middle part". Three different situation for lattice vibrations
are considered : (i) the amplitude of all the atom vibrations in the film is
the same to that in the bulk case, estimated from Eq.~\ref{amplitude1}; (ii)
the surface atomic vibrations are decomposed into the in-plane surface vibration
and the perpendicular component, and their amplitudes are taken from former
experimental results\cite{G.Beni,D.E.Fowler,G.A.Stewart}; the atom vibrations in
the middle part is the same to the bulk case; (iii) freezing the surface atoms in
their equilibrium positions, and treating the atoms in the middle part the same in bulk case.
\begin{figure}
\center
\includegraphics[scale=0.6]{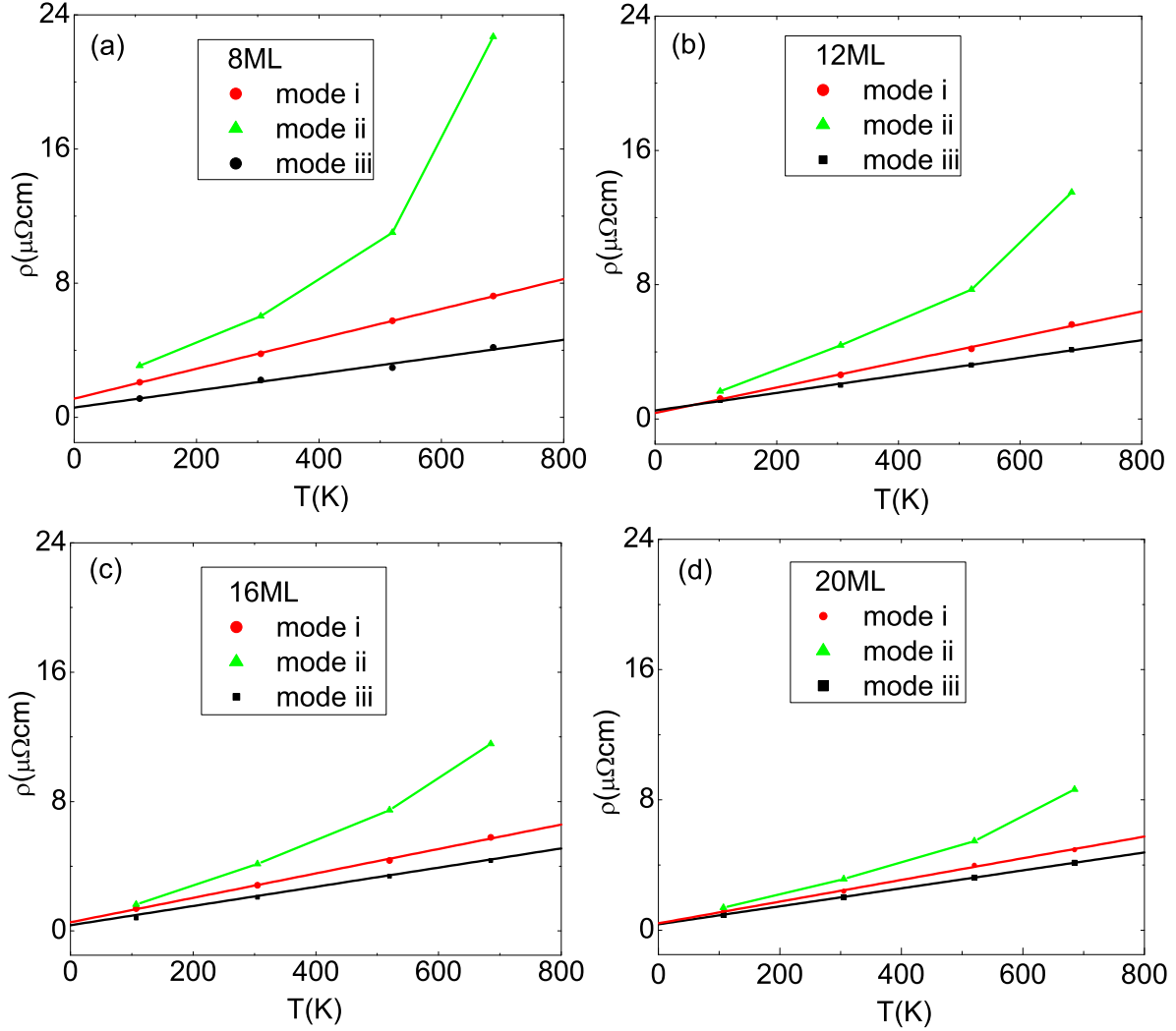}
\caption{(Color online) The resistivity of Cu films with different
thickness. The results for mode (i), mode (ii) and mode (iii) are
represented by red, green, and black lines, respectively.
}\label{3mode}
\end{figure}

The calculated results are presented in Fig.~\ref{3mode}. It shows
that the resistivity increases linearly with temperature for mode
(i) and (iii), which is the same to that for the bulk one. The
resistivity is slightly lower in mode (iii) than mode (i) due to the
freezing of surface atoms. For mode (ii), the two components are
unequal, and the ratios of the surface to bulk vibrational
amplitudes are 1.71 ($A_{\parallel}$) and 1.53 ($A_{\perp}$) when
$T=107~{\rm K}$. When the temperature is $685~{\rm K}$, they
increase to $2.76$ ($A_{\parallel}$) and $2.02$ ($A_{\perp}$),
respectively.\cite{D.E.Fowler} Fig.~\ref{3mode} shows that the
resistivities (the green lines) are largely increased when the
surface vibration is considered. Furthermore, the resistivity
increases non-linearly with the increasing of temperature, specially
for $d=8~{\rm ML}$. These phenomena can be understood by the
following arguments. The resistivity is heavily influenced by the
surface atomic vibration, and the surface vibrational amplitudes
used in our calculations are non-linearly depending on the
temperature due to the reduced symmetry\cite{D.E.Fowler}. The
surface atomic vibration dominates the contribution to the
electron-phonon scattering because the surface to bulk ratio of the
atomic numbers is very large for thin films. Therefore one observes
the largest resistivity and the strongest non-linear temperature
dependence in Fig.~\ref{3mode}(a). As the increasing of the
thickness, the ratio of the surface to bulk atomic number reduces
and thus the contribution from the surface vibration shrinks. One
may find that the temperature dependence of the resistivity becomes
close to those of model (i) and (iii) as shown in
Fig.~\ref{3mode}(d).

The computational results mentioned above suggest that the
non-linear temperature dependence of the resistivity might be
observed in very thin films. Actually, it is not easy to obtain very
thin films or wires in experiment. Recently, Plombon \emph{et
al.}\cite{J.J.Plombon} have observed a linear high-temperature
dependence of the resistivity in the copper wires with the
transversal dimensions ranging from 75~nm (about 400~MLs) to 520~nm
(about 2880~MLs). This work implies that, to observe the nonlinear
temperature dependence, more effort should be made to grow thin
films or nanowires with smaller cross section.
\begin{table}[!h]
\tabcolsep 2.2mm \caption{The extra resistivity of Cu films due to
surface electron-phonon scattering (defined by
$\rho_s\equiv\rho_{ii}-\rho_{iii}$). The unit of $\rho_s$ is
$\mu\Omega cm$.}\label{tableamp2}
\begin{center}
\def\temptablewidth{0.42\textwidth}
\begin{ruledtabular}
\begin{tabular}{ccccc}
{$T(K)$}&{$\rho_s(8{\rm
~ML})$}&{$\rho_s(12{\rm
~ML})$}&{$\rho_s(16{\rm
~ML})$}&{$\rho_s(20{\rm ~ML})$}
\\ \hline
107  &1.963  &0.756      &0.599   &0.419\\
305  &3.800  &1.974     &1.716  &1.117\\
520  &8.039  &3.969     &2.875  &2.233\\
685  &18.52 &8.705     &6.209  &4.489\\
\end{tabular}
\end{ruledtabular}
\end{center}
\end{table}

In order to illustrate more clearly the effect of the surface
electron-phonon scattering, we list the extra resistivity $\rho_{s}$
in table \ref{tableamp2}. The extra resistivity is defined by
$\rho_{s}\equiv\rho_{ii}-\rho_{iii}$, where $\rho_{ii}$ and
$\rho_{iii}$ are the resistivities of Cu film for mode (ii) and mode
(iii), respectively. It is the measure of the contribution due to
the surface electron-phonon scattering. In the temperature range
from $107~{\rm K}$ to $685~{\rm K}$, the extra resistivity varies
from $0.4192\sim18.5215 \mu\Omega cm$ for the Cu films of different
thicknesses. It is quite obvious that the surface electron-phonon
scattering will become more and more important in the resistivity of
Cu films when the temperature increases. This result is consistent
with the experimental observation of Plombon {\emph{et
al.}}.\cite{J.J.Plombon} Their result showed that, for the copper
wires with transverse dimension of 75~nm, the contribution from
electron-phonon scattering increases from 16\% to 63\% when the
temperature changes from $T=20~{\rm K}$ to $T=300~{\rm K}$.
Furthermore, we would also like to compare our result with the
surface roughness contribution of the thin Cu films observed by Ke
\emph{et al.}\cite{Youqi Ke} They showed that the surface roughness
dependent resistivity are about $2\sim14~{\rm \mu}\Omega {\rm cm}$,
which falls in the above mentioned range of the extra resistivity.
Thus one may conclude that the surface electron-phonon scattering
plays a role at least as important as the surface roughness in the
enhanced resistivity of Cu films.

\begin{figure}
\center
\includegraphics[width=3.4in]{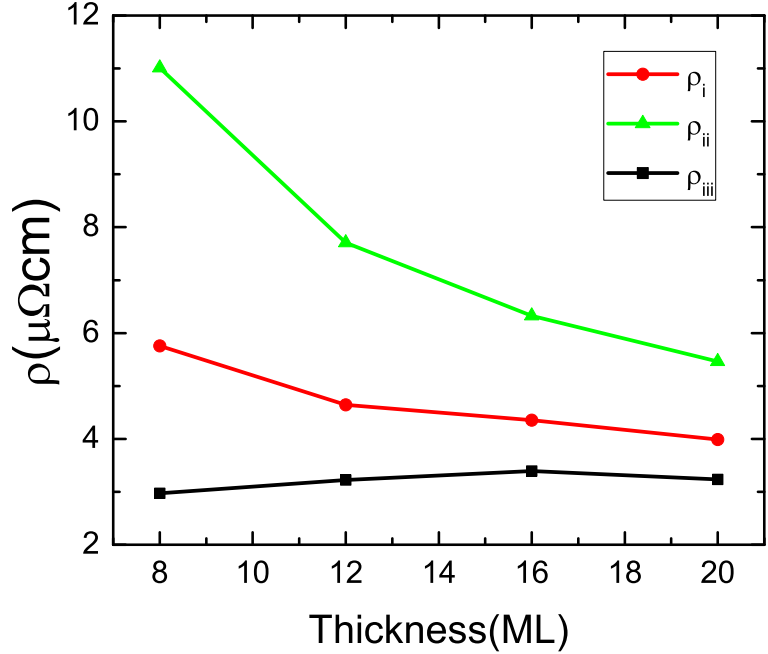}
\caption{(Color online) The resistivity of Cu films as a function of
thickness when the temperature is 520K.}
\end{figure}

It can also be found in table \ref{tableamp2} that the extra
resistivity $\rho_{s}$ decreases with the increase of the film
thickness, which obviously shows the ``size effect". To make it
clear, plotted in Fig.~4 is the thickness dependence of the
resistivities for different models at $T=520~{\rm K}$. One may find
that the value of $\rho_{iii}$ keeps almost unchanged for different
thicknesses and even shows a slightly decrease as the thickness
decreases, while $\rho_{i}$ and $\rho_{ii}$ increase with the
decrease of the thickness. These results reveal that the surface
atomic vibration dominates the contribution to the ``size effect" of
the overall resistivity.

Actually, the electron momentum can be decomposed into two
components, i.e. the ones that are perpendicular and
parallel to the surface, respectively. But only the perpendicular
component affects the resistivity, which is called ``specular
electron scattering". Chawla and Gall\cite{J.S.Chawla} have showed
that it is possible to reduce the surface scattering by realizing
the specular electron scattering at single-crystal Cu surface. As we
have presented in the previous sections, model (iii) shows the
smallest resistivity and weak size effect due to its specular
surface feature. Here, we may suggest that freezing the surface
vibration is also necessary for practical applications to reduce the
resistivity induced by the surface electron-phonon scattering.

\section{Summary}\label{section4}

In summary, we have estimated the mean-square amplitude of atomic
vibration based upon the Debye model. The mean-square amplitude was
used to simulate the atomic displacements during lattice vibration
in the bulk and film samples of Cu. The temperature dependence of
the resistivity for bulk Cu was calculated and the result agrees
reasonably well with the experiment one. The resistivity for three
different Cu film models and various thicknesses were calculated and
analyzed. The result shows that the surface electron-phonon
scattering plays a key role in the enhancement of the resistivity at
high temperature, especially for the thinner Cu films due to the quantum
size effect. Comparing the conductivities for three surface
vibration modes, we may suggest that freezing the surface vibration
is necessary for practical applications to reduce the resistivity
due to the surface electron-phonon scattering.

\acknowledgments {Y.N. Zhao thanks Y. Wang for useful discussions.
This work was supported by the Key Project of Chinese Ministry of
Education (Grant No. 108118). K. Xia acknowledges the support of
National Basic Research Program of China (973 Program).}

\end{document}